\documentclass[10pt, twocolumn]{IEEEtran}

\usepackage{amsmath}
\usepackage{amsfonts}
\usepackage{amssymb}
\usepackage{algorithm}
\usepackage{algorithmic}
\usepackage{graphicx}
\usepackage{verbatim}

\newtheorem{example}{Example}

\newcommand{\beq}{\begin{equation}}
\newcommand{\eeq}{\end{equation}}
\newcommand{\beqnn}{\begin{equation*}}
\newcommand{\eeqnn}{\end{equation*}}
\newcommand{\beqy}{\begin{eqnarray}}
\newcommand{\eeqy}{\end{eqnarray}}
\newcommand{\beqynn}{\begin{eqnarray*}}
\newcommand{\eeqynn}{\end{eqnarray*}}
\newcommand{\bit}{\begin{itemize}}
\newcommand{\eit}{\end{itemize}}
\newcommand{\ben}{\begin{enumerate}}
\newcommand{\een}{\end{enumerate}}
\newcommand{\bex}{\begin{example}}
\newcommand{\eex}{\end{example}}

\newcommand{\balg}[1]{\begin{algorithm} \caption{#1}}
\newcommand{\ealg}{\end{algorithm}}

\newcommand{\balgc}{\begin{algorithmic}[1]}
\newcommand{\ealgc}{\end{algorithmic}}

\newcommand{\bary}{\begin{array}}
\newcommand{\eary}{\end{array}}
\newcommand{\bmx}{\begin{bmatrix}}
\newcommand{\emx}{\end{bmatrix}}
\newcommand{\bsmx}{\left[\begin{smallmatrix}}
\newcommand{\esmx}{\end{smallmatrix}\right]}
\newcommand{\bmxc}[1]{\left[\begin{array}{@{}#1@{}}}
\newcommand{\emxc}{\end{array}\right]}
\newcommand{\bcn}{\begin{center}}
\newcommand{\ecn}{\end{center}}

\newcommand{\s}{\boldsymbol{s}}

\usepackage[rgb]{xcolor}

\usepackage{psfrag}
\usepackage{amssymb}
\usepackage{amsmath}
\usepackage{cite}
\usepackage{graphics}
\usepackage{graphicx}
\usepackage{epsfig}
\usepackage{subfigure}
\usepackage{url}
\usepackage{amscd}
\usepackage{threeparttable}
\usepackage{enumerate}

\begin{document}

\title{An Actor-Critic Reinforcement Learning Method for Computation Offloading with Delay Constraints in Mobile Edge Computing}
\author{Qizhen~Li
\thanks {Qizhen Li is with the School of Information Science and Technology, Southwest Jiaotong University, Chengdu 611756, China (email: liqizhen@my.swjtu.edu.cn).}
}

\maketitle

\bibliographystyle{unsrt}
\thispagestyle{plain}\pagestyle{plain}

\begin{abstract}
In this paper, we consider a mobile edge computing system that provides computing services by cloud server and edge server collaboratively. The mobile edge computing can both reduce service delay and ease the load on the core network. We model the problem of maximizing the average system revenues with the average delay constraints for different priority service as a constrained semi-Markov decision process (SMDP). We propose an actor-critic algorithm with eligibility traces to solve the constrained SMDP. We use neural networks to train the policy parameters and the state value function's parameters to continuously improve the system performance.
\end{abstract}


\section{Introduction}
The most important function of mobile edge computing is providing low delay service for mobile terminals \cite{Hu2015mobile}. In \cite{Li2018SMDP},we proposed maximizing the system revenues using reinforcement learning in cloud-fog computing systems. However, we did not consider delay constraints and wireless bandwidth resources. In this paper, we solve the system revenue optimization problem with delay constraints using a multi-timescale actor-critic reinforcement learning. Compared with the original constrained actor-critic algorithm \cite{Bhatnagar2012online}, we use neural networks to approximate parameterized policy and parameterized state value function to avoid finding proper feature functions. In addition, we use eligibility trace in both actor and critic. In the future, we will present experiment results and more technical details.

\section{System Model}
\begin{figure}[htbp]
    \centering
    \includegraphics[width=60mm,height=70mm]{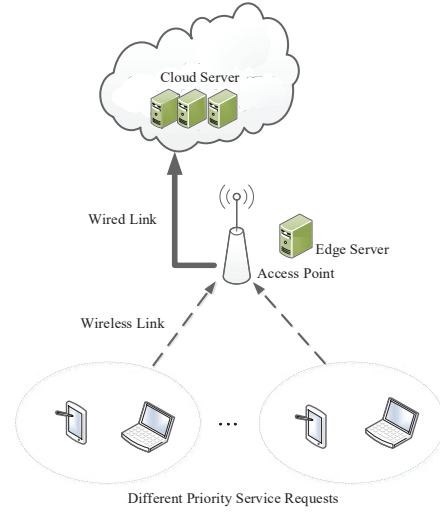}\\
    \caption{System model of mobile edge computing.}\label{fig:Model_MEC}
\end{figure}
We consider a mobile edge computing system, as shown in Fig. \ref{fig:Model_MEC}. Two computing servers are considered: cloud server and edge server. The edge server is in an access point (AP). The access point connects the cloud server and user terminals through wired and wireless links, respectively. The computing resources are quantified as virtual machines (VMs). The wireless bandwidth is quantified as subchannels. We assume that the cloud server has sufficient computing resources and the number of VMs in the edge server is $M$. We assume that the number of subchannels of the wireless link is $B$. We assume that the user terminals have $P$ priority services. Different priority services require different numbers of VMs and subchannels, and different delay constraints.

The considered mobile edge computing system is an event-triggered decision making system. The decision maker is in the AP. When a service request arrives, the decision maker decides whether to accept it, where to process it, and how many VMs and subchannels to be allocated for it. When a service completes, the decision maker spares the VMs and subchannels and transmits the processed data to the user terminal. The objective of the decision making system is to optimize the average system revenues to guarantee the average delay constraints of different priority services.

\section{Constrained SMDP Formulation}
We use a constrained semi-Markov decision process (SMDP) to model the constrained optimization problem mentioned in the previous section. In general, a constrained SMDP can be formulated as a 6-tuple $\{t_i,\mathcal{S},\mathcal{A},q,r, g\}$. Here, $t_i$ is the $i$th decision epoch, $\mathcal{S}$ is the state space, $\mathcal{A}$ is the action space, $q$ is the transition probability, $r$ is the immediate reward of the objective, $g$ is the immediate cost vector of constraints. In the rest of this section, we construct the detailed constrained SMDP for the considered problem.

\subsection{Decision Epoch}
A mentioned in the previous section, the mobile edge computing system is event-triggered. We define the decision epoch is the time instant when a service request arrival or departure occurs.

\subsection{State Space and Action Space}
The total number of $p$th priority ongoing services that occupy $i$ subchannels and $j$ VMs in the edge server is $x_{i,j}^p$. The number of $p$th priority ongoing services occupying $i$ sunchannels in cloud server is $y_i^p$. We assume the maximal number of VMs that a service can occupy is $m$, where $m \leq M$. We also assume that the maximal number of subchannels that a service can occupy is $b$, where $b\leq B$. We assume the cloud server allocates each service sufficient VMs, such as $m$. $x_{i,j}^p$ and $y_i^p$ have to satisfy the following constraints:
\begin{equation}\label{}
  \sum_{i=1}^{b}\sum_{j=1}^{m}\sum_{p=1}^P (ix_{i,j}^p)+\sum_{i=1}^{b}\sum_{p=1}^P (iy_i^p)\leq B
\end{equation}
\begin{equation}\label{}
  \sum_{i=1}^{b}\sum_{j=1}^{m}\sum_{p=1}^P (jx_{i,j}^p)\leq M
\end{equation}

We define an event $e(s)\in \{A_p,D_i^j,F_i\}$. Here, $A_p$ represents the arrival event of the $p$th priority service request, $D_i^j$ represents the departure event of the service which occupies $i$ subchannels and $j$ VMs in the edge server, $F_i$ represents the departure event of the service occupying $j$ subchannels in the cloud server.

We only consider the state in the decision epoch. The state space is $\mathcal{S}=[x_{1,1}^1,\ldots,x_{b,m}^P,y_1^1,\ldots,y_{b}^{P},e(s)]$. In our considered constrained SMDP problem, the decision maker make a decision only in service request. When the service departs, the decision maker naturally releases the bandwidth and computing resources.

The action space is $\mathcal{A}=\bigcup_{s\in \mathcal{S}}\mathcal{A}_s:\{-1,0,a_{i,j}^p,a_i^p\}~(i=\{1,2,\ldots,b\},j\in\{1,2,\ldots,m\},p\in\{1,2,\ldots,P\})$. Here, $-1$ represents releasing the computing and bandwidth resources when a service departs, $0$ represents rejecting a service request, $a_{i,j}^p$ represents an action that edge server accepts the $p$th priority service request and allocates $i$ subchannels and $j$ VMs, $a_i^p$ represents an action that cloud server accepts the $p$th priority service request and allocates $i$ subchannels.

\subsection{Transition Probability}
We define the transition probability is $q(t,s'|s,a)$, where $s'$ is the next state. The state transition probability is
\begin{equation}\label{}
  \Pr(s'|s,a)=\int_{t=0}^{\infty}dq(t,s'|s,a).
\end{equation}
The expected time interval between adjacent decision epochs is defined as $\tau(s,a)$, whose formulation is as follows:
\begin{equation}\label{}
  \tau(s,a)=\sum_{\s'\in\mathcal{S}}\int_{t=0}^{\infty}tdq(t,s'|s,a)
\end{equation}

\subsection{Immediate Reward and Constraints}
The immediate reward we consider is the net reward which consists of the reward got at the decision epoch and the cost lost between the last two decision epochs. The immediate reward is formulated as follows:
\begin{equation}\label{}
\begin{split}
    r(s,a,s')=
    \begin{cases}
        k_c- c(s,a)\tau(s,a,s'),~a=a_i^p \\
        k_e- c(s,a)\tau(s,a,s'),~a=a_{i,j}^p\\
        -k_r- c(s,a)\tau(s,a,s'),~a=0\\
        -c(s,a)\tau(s,a,s'),~a=-1,
    \end{cases}
\end{split}
\end{equation}
where $k_c$ and $k_e$ represent the reward that the service request is accepted by the cloud server and edge server, respectively. $k_r$ is the penalty of rejecting a service request. $\tau(s,a,s')$ is time interval between the state $s$ and $s'$. $c(s,a)$ is the loss ratio after taking action $a$ at state $s$, as follows:
\begin{equation}\label{}
  c(s,a)=c_c\sum_{i=1}^{b}\sum_{j=1}^{m}\sum_{p=1}^P (jx_{i,j}^p)+c_e\sum_{i=1}^{b}\sum_{p=1}^P (my_i^p)
\end{equation}
where $c_c$ and $c_e$ represent the cost of running a VM per unit time in cloud server and edge server, respectively.

According to the Little`s Law, we can use the length of the queue to represent the delay \cite{Comaniciu2003jointly}. The immediate constraint of $p$ priority service as follows:
\begin{equation}\label{}
  g_p(s,a,s')=\sum_{i=1}^{b}\sum_{j=1}^{m}w_i^jx_{i,j}^p\tau(s,a,s')+\sum_{i=1}^{b}w_iy_j^p\tau(s,a,s'),
\end{equation}
where $w_i^j$ and $w_i$ represent the weights.

\subsection{Policy Optimization with Delay Constraints}
We define the policy $\pi$ as a map from a state to an action. From an initial state $s_0$, the objective of this paper is to find an optimal policy to satisfy the following optimization problem:
\begin{equation}\label{CP}
  \max_{\pi}~ J(\pi)\doteq\lim_{N\rightarrow \infty} \frac{\textrm{E}_{s_0}^{\pi}\left[\sum_{n=0}^{N}r(s_n,a_n)\right]}{\textrm{E}_{s_0}^{\pi}\left[\sum_{n=0}^N\tau_n\right]}
\end{equation}
\begin{equation*}
\begin{split}
  &\textrm{s.b.}~ G_p(\pi)\doteq\lim_{N\rightarrow \infty} \frac{\textrm{E}_{s_0}^{\pi}\left[\sum_{n=0}^{N}g_p(s_n,a_n)\right]}{\textrm{E}_{s_0}^{\pi}\left[\sum_{n=0}^N\tau_n\right]},\\
  &~~~~~~p=1,2,\ldots,P.
  \end{split}
\end{equation*}
We consider the stationary probability of state $s$ following policy $\pi$ as $d^{\pi}(s)$. In a constrained MDP, the optimal policy is always random. We set the probability taking action $a$ in state $s$ as $\pi(a|s)$. Thus, the constrained optimization problem \eqref{CP} can be formulated as:
\begin{equation}\label{CP2}
  \max_{\pi}~ J(\pi)\doteq\sum_{s\in\mathcal{S}}d^{\pi}(s)\sum_{a\in\mathcal{A}_s}\pi(a|s)r(s,a)
\end{equation}
\begin{equation*}
\begin{split}
  &\textrm{s.b.}~ G_p(\pi)\doteq\sum_{s\in\mathcal{S}}d^{\pi}(s)\sum_{a\in\mathcal{A}_s}\pi(a|s)g_p(s,a), \\
  &~~~~~~p=1,2,\ldots,P.
\end{split}
\end{equation*}

\section{Actor-Critic Algorithm for Constrained SMDP}
We formulate the constrained optimization problem \eqref{CP2} as the following Lagrangian:
\begin{equation}\label{Lagrangian}
\begin{split}
    & L(\pi,\boldsymbol{\gamma}) \doteq J(\pi)+\sum_p \gamma_p(\alpha_p-G_p(\pi)) \\
    & = \sum_{s\in\mathcal{S}}d^{\pi}(s)\sum_{a\in\mathcal{A}_s}\pi(a|s)\left[r(s,a)-\sum_p\gamma_p(g_p(s,a)-\alpha_p)\right]
\end{split}
\end{equation}
where $\boldsymbol{\gamma}=[\gamma_1,\gamma_2\ldots\gamma_P]^T$ is the vector of lagrange multiplier, $\gamma_p\in \mathbb{R}^{+} \bigcup \{0\}$.

The objective of the Lagrangian is to find $\pi$ and $\boldsymbol{\gamma}$ to satisfy the following expression:
\begin{equation}\label{}
  \max_{\pi}\min_{\boldsymbol{\gamma}} L(\pi,\boldsymbol{\gamma}).
\end{equation}

Given the Lagrange multiplier $\gamma$, the optimal policy $\pi_{\gamma}^*$ satisfies the following Bellman equation:
\begin{equation}\label{}
\begin{split}
  &\beta^{*,\gamma}+v^{*,\gamma}(s)=\max_{a\in\mathcal{A}_s}\left(r(s,a)-\sum_{p=1}^P\gamma_p(g_p(s,a)-\alpha_p)\right. \\
  &~~~~~~~~~~~~~~~~~~~~~~~~~~\left.+\sum_{s'\in S}\Pr(s'|s,a)v^{*,\gamma}(s')\right).
\end{split}
\end{equation}
According to Poisson equation \cite{Bhatnagar2012online}, for any $\pi$ and $\gamma$, the following equation satisfies:
\begin{equation}\label{}
  \begin{split}
  &\beta^{\pi,\gamma}+v^{\pi,\gamma}(s)=\sum_{a\in\mathcal{A}_s}\pi(a|s)\left(r(s,a)-\sum_{p=1}^P\gamma_p(g_p(s,a)-\alpha_p)\right. \\
  &~~~~~~~~~~~~~~~~~~~~~~~~~~\left.+\sum_{s'\in S}\Pr(s'|s,a)v^{\pi,\gamma}(s')\right).
\end{split}
\end{equation}

We use actor-critic reinforcement learning algorithm to find the optimal policy $\pi^*$. Firstly, we respectively parameterize the random policy and state value function as follows:
\begin{equation}\label{}
  \hat{\pi}(a|s;\boldsymbol{\theta})\approx \pi(a|s),
\end{equation}
\begin{equation}\label{}
  \hat{v}(s,\textbf{w})\approx v(s).
\end{equation}
We use soft-max in action preferences for policy parameterization. We set parameterized numerical preferences to $h(s,a, \boldsymbol{\theta})\in \mathbb{R}$. Parameterized policy can be formulated as
\begin{equation}\label{}
  \hat{\pi}(a|s;\boldsymbol{\theta})\doteq \frac{e^{h(s,a, \boldsymbol{\theta})}}{\sum_be^{h(s,b, \boldsymbol{\theta})}}.
\end{equation}
We use neural networks to represent $h(s,a, \boldsymbol{\theta})$ and $\hat{v}(s,\textbf{w})$.

We use multi-timescale stochastic approximation (MTSA) algorithm to find optimal $\boldsymbol{\theta}$, $\textbf{w}$ and $\gamma$. We set multi-timescale step-sizes as $a(n)$, $b(n)$, $c(n)$ and $d(n)$ satisfying the following conditions:
\[\sum_n a(n)=\sum_n b(n)=\sum_n c(n)=\infty,\]
\[\sum_n(a^2(n)+b^2(n)+c^2(n))<\infty,\]
\[\lim_{n\rightarrow\infty}\frac{b(n)}{a(n)}=\lim_{n\rightarrow\infty}\frac{c(n)}{b(n)} =0,\]
\[d(n)=Ca(n),\]
where $C$ is a positive constant.

We set $Y_p(n)$ is the estimate of $G_p(\boldsymbol{\theta})$ in the $n$th step. We set the delay trace rates as $\lambda^{\textbf{w}}\in [0,1]$ and $\lambda^{\boldsymbol{\theta}}\in [0,1]$. We initialize the average Lagrange reward as $\bar{R}(0) =0$.

Loop:
\[\delta(n)\leftarrow r(n)-\sum_{p=1}^P\gamma_p(g_p(n)-\alpha_p)-\bar{R}(n)+\hat{v}(s',\textbf{w})-\hat{v}(s,\textbf{w}),\]
\[\bar{R}(n+1)\leftarrow \bar{R}(n) +d(n)\delta,\]
\[\textbf{z}^{\textbf{w}}\leftarrow \lambda^{\textbf{w}}\textbf{z}^{\textbf{w}} +\nabla\hat{v}(s,\textbf{w}),\]
\[\textbf{z}^{\boldsymbol{\theta}}\leftarrow \lambda^{\boldsymbol{\theta}}\textbf{z}^{\boldsymbol{\theta}}+\nabla \ln\hat{\pi}(a|s,\boldsymbol{\theta}),\]
\[\textbf{w}\leftarrow \textbf{w}+a(n)\delta(n)\textbf{z}^{\textbf{w}},\]
\[\boldsymbol{\theta}\leftarrow \Gamma\left[\boldsymbol{\theta}+b(n)\delta(n)\textbf{z}^{\boldsymbol{\theta}}\right],\]
\[\gamma_p(n+1)=\hat{\Gamma}\left[\gamma_p(n)+c(n)(Y_p(n)-\alpha_p))\right],\]
\[Y_p(n+1)=Y_p(n)+a(n)(g_p(n)-Y_p(n)),\]
where $\Gamma$ and $\hat{\Gamma}$ are to guarantee policy parameters and Lagrange multipliers in the feasible region.

\end{document}